\documentclass[10pt,aps,pra,showpacs,twocolumn,superscriptaddress]{revtex4-1}
\usepackage{graphicx}
\usepackage[usenames]{color}
\usepackage{amssymb}
\usepackage{amsmath}
\newcommand{\eq}[1]{Eq.~(\ref{#1})}
\newcommand{\fig}[1]{Fig.~\ref{#1}}
\newcommand{\be}[1]{\begin{equation}\label{#1}}
\newcommand{\ee}{\end{equation}}

\begin{document}

\title{ Microcanonical distribution for one-electron triatomic molecules}
\author{C. Lazarou}
\address{Department of Physics and Astronomy, University College London, Gower Street, London WC1E 6BT, United Kingdom}
\author{A. Chen}
\address{Department of Physics and Astronomy, University College London, Gower Street, London WC1E 6BT, United Kingdom}
\author{A. Emmanouilidou}
\address{Department of Physics and Astronomy, University College London, Gower Street, London WC1E 6BT, United Kingdom}

\begin{abstract}
We formulate a microcanonical distribution for an arbitrary one-electron triatomic molecule. This distribution can be used to describe the initial state in strongly-driven two-electron triatomic molecules. Namely, in many semiclassical models 
that describe ionization of  two-electron molecules driven by intense infrared laser fields in the tunneling regime initially one electron tunnels while the other electron is bound. The microcanonical distribution presented in this work  
can be used to describe the initial state of this bound electron.
\end{abstract}
\pacs{33.80.Rv, 34.80.Gs, 42.50.Hz}
\date{\today}
%
\maketitle

The nonlinear response of multi-center molecules to intense laser fields is a fundamental problem. For instance, 
understanding the break-up dynamics of strongly-driven molecules paves the way for controlling and imaging molecular processes \cite{Imaging}. 
Semi-classical models are essential in understanding the dynamics during the fragmentation of multi-center molecules driven by intense infrared laser pulses. One reason is 
that treating the dynamics of electrons and nuclei at the same time poses an immense challenge for fully ab-initio quantum mechanical calculations. Currently quantum mechanical techniques can only address one electron in triatomic molecules in two-dimensions \cite{Bandrauk}. Semi-classical models  have provided significant insights, for example, in double ionization of strongly-driven  $\mathrm{H_{2}}$ with ``fixed nuclei" \cite{Chinesemolecule, Emmanouilidou2009} and  in  ``frustrated" double ionization during the fragmentation of  strongly-driven $\mathrm{H_{2}}$ \cite{Emmanouilidou2012, Emmanouilidou2014}, where one electron eventually stays bound in a highly excited state of  the H atom.

The initial state that is commonly employed by semi-classical models  for strongly-driven two-electron atoms and molecules, for intensities in the tunneling regime,  involves one electron that tunnel-ionizes in the field-lowered Coulomb potential and another electron that remains bound. The electron that tunnel-ionizes emerges from the barrier with a zero velocity along the direction of the laser field, while its velocity perpendicular to the laser field is given by a Gaussian distribution \cite{ADK}. The tunneling rate can be obtained using semi-classical treatments, for instance see  \cite{ADK} for atoms and  \cite{Murray2011} for molecules.  The electron that is initially bound is commonly described in semi-classical models by a microcanonical distribution.  To our knowledge, in the literature, a microcanonical distribution is available only for diatomic molecules  \cite{Olson}. In this work, we formulate a microcanonical distribution for  any one-electron triatomic molecule which can also describe the initial state of the bound electron in the above described semi-classical models.

\section{Microcanonical distribution}
In the following, we  formulate a one-electron microcanonical distribution for triatomic molecules. We denote the positions of the nuclei by $\mathrm{{\bf R}_a=(0,0,-R_{a b}/2)}$, $\mathrm{{\bf R}_{b}=(0,0,R_{a b}/2)}$ and $\mathrm{{\bf R}_c=(x_{c},0,z_{c})}$ and the inter-nuclear distances by $\mathrm{R_{a b}}$, $\mathrm{R_{a c}}$ and $\mathrm{R_{b c}}$, see \fig{fig1}. One can show that the coordinates of the nucleus C are expressed in terms of the inter-nuclear distances as follows:

\begin{figure} [ht]
\centering
 \includegraphics[clip,width=0.25\textwidth]{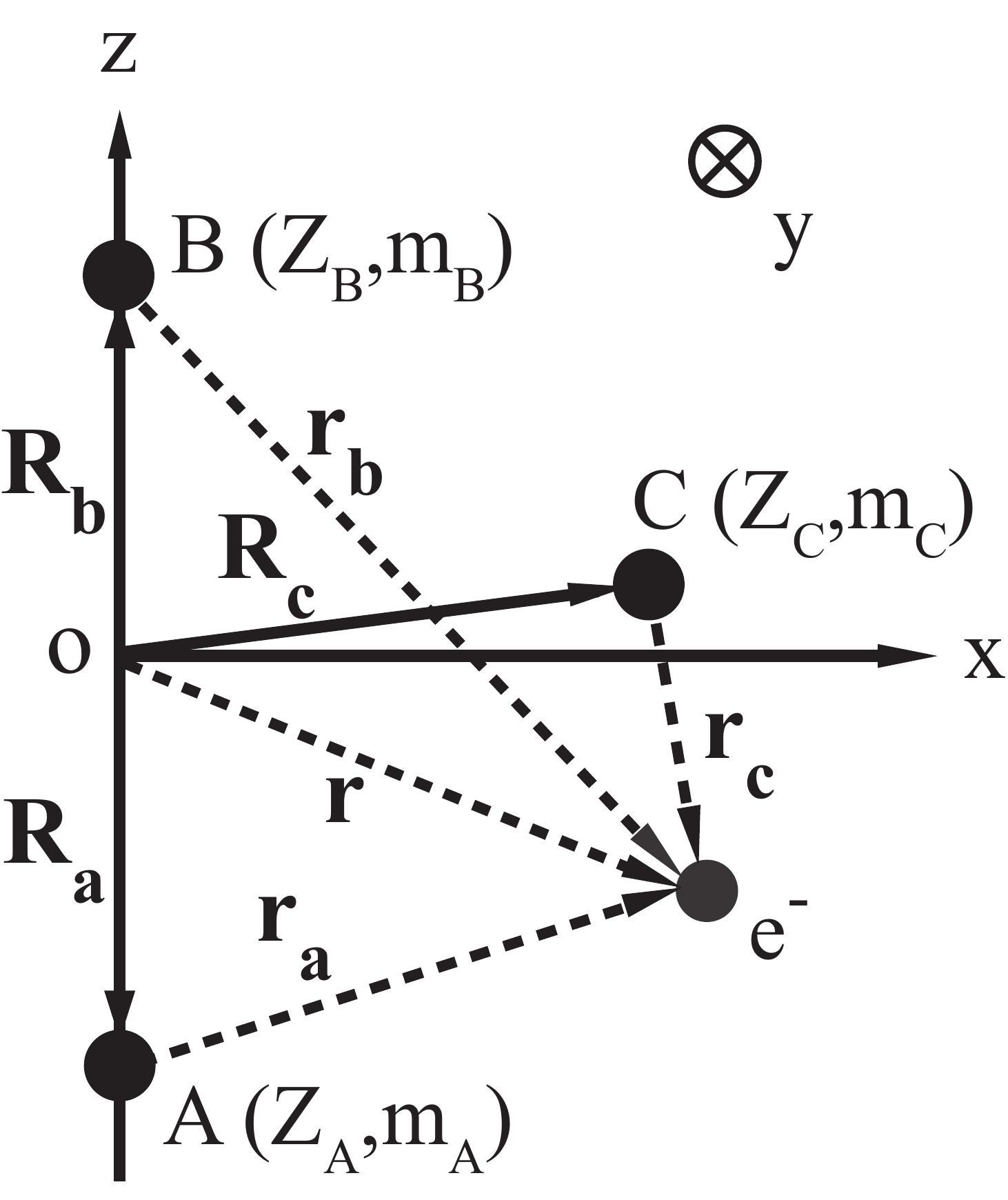}
\caption{The configuration of the triatomic molecule we use to set-up  the microcanonical distribution.}
\label{fig1}
\end{figure}

\begin{equation}
\mathrm{z_{c}=\frac{R_{a c}^2-R_{bc}^2}{2 R_{a b}}}, \hspace{6pt}\mathrm{x_{c}=\pm\sqrt{R_{a c}^2-\left(\frac{R_{a c}^2-R_{bc}^2+R_{a b}^2}{2 R_{a b}}\right)^2}}.
\end{equation} 
 We denote the position vector of the electron by $\mathrm{{\bf r}}$ and the distances of the electron from nuclei A, B and C  by $\mathrm{r_{a}=|{\bf r}-{\bf R}_a|}$,  $\mathrm{r_{b}=|{\bf r}-{\bf R}_b|}$ and $\mathrm{r_{c}=|{\bf r}-{\bf R}_c|}$, respectively. We then define the confocal elliptical coordinates $\mathrm{\lambda}$ and $\mathrm{\mu}$ using  the nuclei A and B as the foci of the ellipse, that is, 
 
 \begin{equation}
 \mathrm{\lambda=\frac{r_{a}+r_b}{R_{a b}}},  \hspace{6pt}\mathrm{\mu=\frac{r_{a}-r_b}{R_{a b}}},
 \end{equation}
 where $\mathrm{\lambda \in [1,\infty)}$ and $\mathrm{\mu \in [-1,1]}$. The third coordinate  $\mathrm{\phi \in [0,2 \pi]}$ is the angle between the projection of the position vector $\mathrm{{\bf r}}$ on the xy plane and the positive x axis; it thus defines the rotation angle around the axis that passes through nuclei A and B.
 The potential of the electron  in the presence of the nuclei A, B and C, which have  charges $\mathrm{Z_{A}}$, $\mathrm{Z_{B}}$ and $\mathrm{Z_{C}}$, respectively, is given by 
 \begin{equation}
\mathrm{W(r_{a},r_{b},r_c)=-\frac{Z_A}{r_a}-\frac{Z_B}{r_b}-\frac{Z_C}{r_c}}.
\end{equation}
This potential is expressed   in terms of the confocal elliptical coordinates   as follows

\begin{widetext}
\begin{eqnarray}
\mathrm{W(\lambda,\mu, \phi)}&&\mathrm{=-\frac{2}{R_{a b}}\left[\frac{Z_A}{\lambda+\mu} +\frac{Z_B}{\lambda-\mu} +\right.}\nonumber\\
&&\mathrm{\left.Z_{C}\left ( (\lambda^2+\mu^2-1)-\frac{4z_c}{R_{a b}}\lambda \mu-\frac{4x_{c}}{R_{a b}}\cos(\phi)\sqrt{(\lambda^2-1)(1-\mu^2)}+\frac{4(x_c^2+z_{c}^2)}{R_{a b}^2}\right)^{-\frac{1}{2}}\right ]}.
\end{eqnarray}
\end{widetext}
The one-electron microcanonical distribution is given by
\begin{equation}
\mathrm{f({\bf r}, {\bf p})}\propto \mathrm{\delta(E_{i}-\frac{p^2}{2}-W)},
\end{equation}
where $\mathrm{E_{i}=-I_p}$ is the ionization energy of the one-electron triatomic molecule. Note that the energy is given by $\mathrm{E=p^2/2+W}$. The electron momentum in terms of the confocal elliptical coordinates is expressed as follows

\begin{eqnarray}
\mathrm{p_x}&=&\mathrm{\sqrt{2(E-W(\lambda,\mu, \phi))}\cos(\phi_{p})\sqrt{1-\nu_{p}^2}}, \nonumber\\
\mathrm{p_y}&=&\mathrm{\sqrt{2(E-W(\lambda,\mu, \phi))}\sin(\phi_{p})\sqrt{1-\nu_{p}^2}}, \\
\mathrm{p_z}&=&\mathrm{\sqrt{2(E-W(\lambda,\mu, \phi))}\nu_p},\nonumber
\end{eqnarray}
where $\mathrm{\phi_{p}\in [0,2\pi]}$ and $\mathrm{\nu_{p}\in[-1,1]}$ define the momentum $\mathrm{\bf p}$ in spherical coordinates.
Transforming from $\mathrm{({\bf r}, {\bf p})\rightarrow(\lambda,\mu,\phi,E,\nu_{p},\phi_{p})}$ and integrating $\mathrm{f(\lambda,\mu,\phi,E,\nu_{p},\phi_{p})}$
over $\mathrm{E}\in(-\infty,0)$, $\mathrm{\phi_{p}}$ and $\mathrm{\nu_{p}}$ we find 

\begin{equation}
\mathrm{\rho(\lambda,\mu,\phi)\propto (\lambda^2-\mu^2)\sqrt{2(E_{i}-W(\lambda,\mu,\phi))}}.
\end{equation}
The  $\mathrm{\rho}$ distribution goes to zero and is thus well-behaved when the electron is placed on top of either nucleus A or B. However, when $\mathrm{{\bf r}\rightarrow {\bf R}_{c}}$, i.e., the electron is placed on top
of nucleus C, $\mathrm{\rho(\lambda,\mu,\phi)\rightarrow \infty}$. We eliminate this singularity by introducing an additional transformation. Setting $\mathrm{\lambda=\lambda_{c}=(R_{ac}+R_{bc})/R_{ab}}$, $\mathrm{\phi=0}$ and expanding $\mathrm{\rho(\lambda_{c},\mu,0)}$ around $\mathrm{\mu=\mu_c=(R_{ac}-R_{bc})/R_{ab}}$ we find 
\begin{equation}
\mathrm{\rho(\lambda_c,\mu,0)\propto \frac{1}{|\mu-\mu_c|^{1/2}}},
\label{eq:sing}
\end{equation}
where $\mathrm{\lambda_c}$ and $\mathrm{\mu_c}$ are  the values of $\mathrm{\lambda}$ and $\mathrm{\mu}$, respectively, when the electron is placed on top of the nucleus C. To eliminate the singularity in \eq{eq:sing}, we introduce a new variable $\mathrm{t}$, such that $\mathrm{t^{\gamma}=\mu-\mu_c}$. The new distribution takes the form
\begin{widetext}
\begin{eqnarray}
\tilde{\rho}(\lambda,t,\phi)&\propto&\left\{
\begin{array}{lll}
|t^{\gamma-1}|(\lambda^2-(t^{\gamma}+\mu_{c})^2)\sqrt{P(\lambda,t,\phi)} &\mathrm{for} &P(\lambda,t,\phi)\geq 0\\
& & \nonumber\\
0&\mathrm{for}  &P(\lambda,t,\phi)<0,
\end{array}
\right.\\
&&\\
\mathrm{P(\lambda,t,\phi)}&=&\mathrm{2E_{i}+\frac{4}{R_{a b}}}\mathrm{\left[\frac{Z_A}{\lambda+t^{\gamma}+\mu_c} +\frac{Z_B}{\lambda-t^{\gamma}-\mu_c}+Z_{C}\left((\lambda^2+(t^{\gamma}+\mu_c)^2-1)-\frac{4z_c}{R_{a b}}\lambda (t^{\gamma}+\mu_c)-\right.\right.}\nonumber\\
&&\mathrm{\left.\left.\frac{4x_{c}}{R_{a b}}\cos(\phi)\sqrt{(\lambda^2-1)(1-(t^{\gamma}+\mu_c)^2)}+\frac{4(x_c^2+z_{c}^2)}{R_{ab}^2}\right)^{-\frac{1}{2}}\right]}.\nonumber
\end{eqnarray}


\end{widetext}
Since $\mathrm{\mu \in [-1,1]}$, $\mathrm{t^{\gamma}}$ and $\mathrm{t}$ take both negative and positive values and therefore, if we choose one  $\mathrm{\gamma}$ for all values of $\mathrm{\mu}$, $\mathrm{\gamma}$ must be odd.   Moreover, to avoid the singularity when the electron is placed on top of nucleus C, $\mathrm{\gamma}$ must be such that  $\mathrm{t^{\gamma-1}/t^{\gamma/2}\rightarrow 0}$, i.e., $\mathrm{\gamma \geq 2}$. Combining the above two conditions, yields $\mathrm{\gamma=3,5,7,...}$. The new distribution 
$\mathrm{\tilde{\rho}(\lambda,t,\phi)}$ goes to zero when the electron is placed on top of the nucleus C, i.e., when $\mathrm{\lambda=\lambda_c}$, $\mathrm{t=0}$ and $\mathrm{\phi=0,2\pi}$. 

To set up the initial conditions we find $\mathrm{\lambda_{max}}$ so that $\mathrm{p^2/2=E_{i}-W}>0$ and equivalently 
 $\mathrm{P(\lambda,t,\phi)\geq 0}$. We then find the maximum value $\mathrm{\tilde{\rho}_{max}}$ of the distribution $\mathrm{\tilde{\rho}(\lambda,t,\phi)}$, for the allowed values  of the parameters $\mathrm{\lambda}$, $\mathrm{t}$ and $\mathrm{\phi}$. We next generate the uniform random numbers $\mathrm{\lambda \in [1, \lambda_{max}]}$, $\mathrm{t \in [t_{min},t_{max}]}$, $\mathrm{\phi \in [0, 2\pi]}$ and $\mathrm{\chi \in [0, \tilde{\rho}_{max}]}$, with $\mathrm{t_{min}=-(1+\mu_c)^{1/\gamma}}$ and $\mathrm{t_{max}=(1-\mu_c)^{1/\gamma}}$. If $\mathrm{\tilde{\rho}(\lambda,t,\phi)>\chi}$ then the generated values of $\mathrm{\lambda}$, $\mathrm{t}$ and $\mathrm{\phi}$ are accepted as initial conditions, otherwise, they are rejected and the sampling process starts again.

Following the above described formulation, we obtain the initial conditions of the electron with respect to the origin of the coordinate system. To obtain the initial conditions for the position of the electron with respect to the center of mass of the triatomic molecule, $\mathrm{{\bf r'}}$,  in terms of the ones with respect to the origin, $\mathrm{{\bf r}}$,  we shift the coordinates by $\mathrm{{\bf r'}={\bf r}-{\bf R_{cm}}}$, where $\mathrm{{\bf R_{cm}}}$ is given by $\mathrm{(X_{cm},0,Z_{cm})}$ with

\begin{eqnarray}
X_{cm}&=&\frac{m_{C}x_{c}}{m_{A}+m_{B}+m_{C}},\nonumber\\
&&\\
Z_{cm}&=&\frac{R_{ab}(m_{B}-m_{A})/2+m_{C}z_{c}}{m_{A}+m_{B}+m_{C}}, \nonumber
\end{eqnarray}
with $\mathrm{m_{A}}$, $\mathrm{m_{B}}$ and $\mathrm{m_{C}}$ the masses of the nuclei.

As an example, we next obtain the probability densities of the position and the momentum of the electron that is initially bound in  $\mathrm{H_{3}^{+}}$ when the molecule is driven by an intense infrared laser field. We assume the other electron tunnel-ionizes in the initial state.   We consider the $\mathrm{H_{3}^{+}}$ triatomic molecule in its ground state, where the distance of the nuclei in the equilateral triangle arrangement  is 1.65 a.u. and the first and second ionization energies are  $\mathrm{I_{p_1}}=1.2079$ a.u.  and $\mathrm{I_{p_2}}=1.93$ a.u., respectively. We find the ionization potentials and equilibrium distances of the initial state using MOLPRO, which is a quantum chemistry package \cite{MOLPRO}.  For the microcanonical distribution the relevant ionization energy is $\mathrm{I_p=I_{p_{2}}}$, since $\mathrm{I_{p_1}}$ is associated with the electron that tunnel-ionizes in the initial state.

\begin{figure} [ht!]
\centering
 \includegraphics[clip,width=0.40\textwidth]{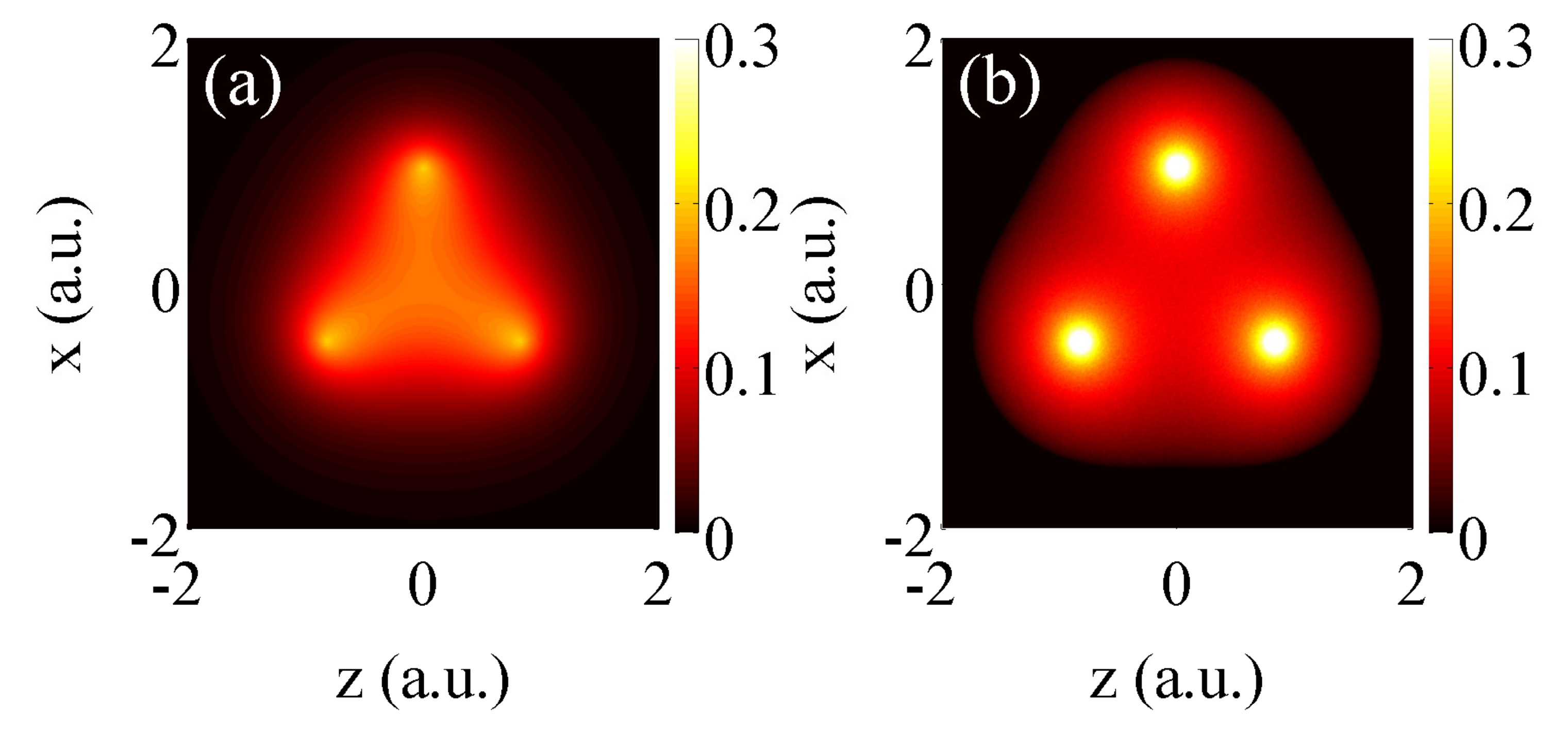}
\caption{The left panel shows the quantum mechanical probability density of the electron position on the x-z plane. The right panel shows the  microcanonical probability density of the electron position on the x-z plane.}
\label{position}
\end{figure}

In \fig{position} (b) we plot the probability density of the position of the electron on the x-z plane for $\mathrm{y=0}$ using the above described microcanonical distribution. To compare, in \fig{position} (a) we plot  the quantum mechanical probability density of the position of the electron on the x-z plane. That is, we plot $\mathrm{|\Psi(x,0,z)|^2}$, where $\mathrm{\Psi({\bf r})}$ is the quantum mechanical wavefunction for the $\mathrm{H_{3}^{2+}}$ molecule, which we obtain using Molpro. The two plots, \fig{position} (a) and (b), show that the two probability densities for the electron position compare well. However, the microcanonical probability density underestimates the electron probability density  between the nuclei and overestimates the electron probability density around the nuclei. 

\begin{figure} [ht!]
\centering
 \includegraphics[clip,width=0.50\textwidth]{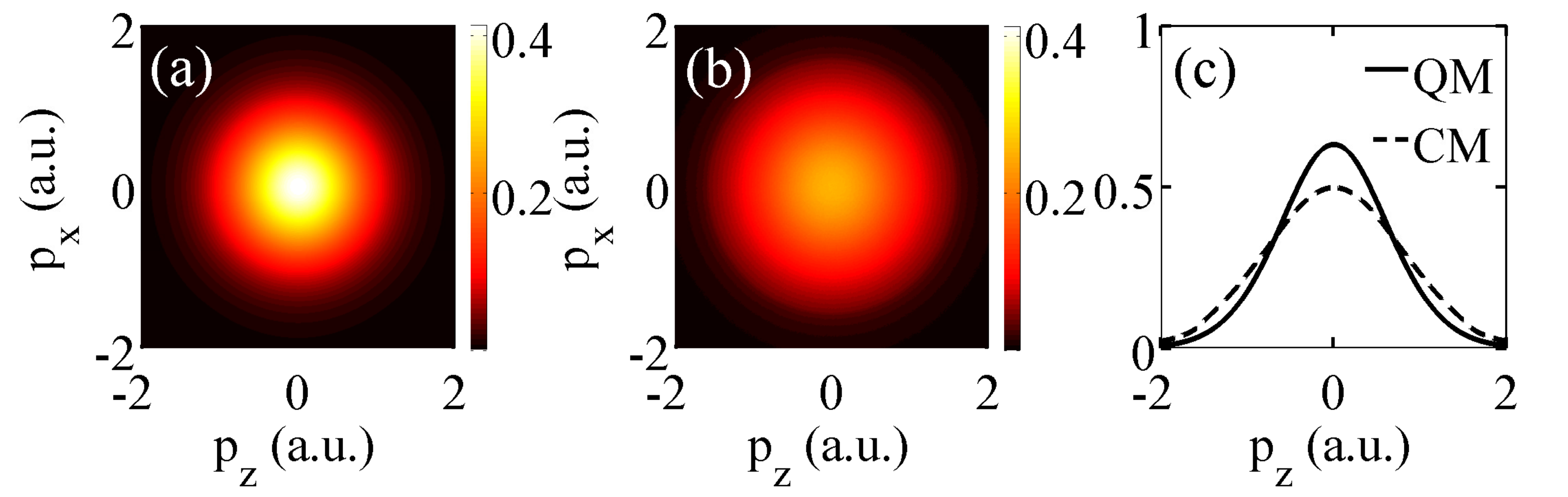}
\caption{The left panel shows the quantum mechanical probability density of the electron momentum  on the $\mathrm{p_{x}-p_{z}}$ plane. The middle panel shows the  microcanonical probability density of the electron momentum plotted on the $\mathrm{p_{x}-p_{z}}$ plane for all values of $\mathrm{p_{y}}$. The right panel shows the projections on the $\mathrm{p_{z}}$ axis of the probability densities plotted in \fig{momentum} (a) and (b). }
\label{momentum}
\end{figure}

In addition, in \fig{momentum} (b) for all values of the electron momentum component along the y-axis, $\mathrm{p_{y}}$, we plot  the probability density  of the electron momentum on the $\mathrm{p_x-p_z}$ plane using the microcanonical distribution.
To compare, in \fig{momentum} (a) we plot  $\mathrm{\rho^{QM}(p_{x},p_{z})}$, which we obtain by first computing the quantum mechanical wavefunction in momentum space using the quantum mechanical wavefunction $\mathrm{\Psi({\bf r})}$ computed  from Molpro  

\begin{eqnarray}
\mathrm{\Phi({\bf{p}})}=\mathrm{\frac{1}{(2\pi)^{3/2}}\int\Psi({\bf{r}})e^{-i{\bf{pr}}}d{\bf{r}}},
\end{eqnarray}
and then by integrating  over $\mathrm{p_{y}}$
\begin{equation}
\mathrm{\rho^{QM}(p_{x},p_{z})=\int_{-\infty}^{\infty}|\Phi({\bf p})|^2dp_{y}}.
\end{equation}
The two plots,   \fig{momentum} (a) and (b), show that the two probability densities for the electron momentum compare well. However, the microcanonical probability density overestimates the higher values of the electron momentum. This can be seen more clearly in \fig{momentum} (c) where we plot the probability density of the electron momentum along the $\mathrm{p_{z}}$ axis both quantum mechanically and using our microcanonical distribution. To obtain the plots in \fig{momentum} (c) we project the probability densities of the electron momentum in \fig{momentum} (a) and (b) on the $\mathrm{p_{z}}$ axis. \fig{momentum} (c) clearly shows that the probability density of the electron momentum obtained from the  microcanonical distribution overestimates the higher values  of the momentum component $\mathrm{p_{z}}$. This is consistent with our previous finding that the microcanonical distribution overestimates values of the electron position around the nuclei resulting to higher values of the momentum. 

In conclusion, in the current work we have formulated a microcanonical distribution for a general one-electron triatomic molecule. This distribution  can be used to describe the initial state of the bound electron in semiclassical models of strongly-driven two-electron triatomic molecules.

{\it Acknowledgments.} A.E. acknowledges the EPSRC grant no. J0171831 and the use of the computational resources of Legion at UCL.


\end{document}